\documentclass[graybox]{svmult}
\usepackage{mathptmx}
\usepackage{amsmath}
\usepackage{microtype}
\usepackage{float}
\usepackage{graphicx}
\usepackage{latexsym}
\usepackage{hyperref} 

\newcommand{\gmunu}{g_{\mu\nu}}

\begin{document}

\title*{INTERACTION OF GRAVITATIONAL WAVES WITH CHARGED PARTICLES}
\titlerunning{Interaction of Gravitational Waves with Charged Particles}
\author{{\bf Thulsi Wickramasinghe, Will Rhodes, Mitchell Revalski}}
\institute{Thulsi Wickramasinghe \and Will Rhodes \and Mitchell Revalski \at The College of New Jersey, 2000 Pennington Road, Ewing, New Jersey, 08628, USA \\ \email{wick@tcnj.edu,  rhodesw1@tcnj.edu, revalsm1@tcnj.edu}}
\maketitle

\vspace {-30 mm} 
{\bf
This is part one of two. For part two see:\\
\href{http://adsabs.harvard.edu/abs/2015ASSP...40..301R}{http://adsabs.harvard.edu/abs/2015ASSP...40..301R}
}\\

\abstract{
It is shown here that a cloud of charged particles could in principle absorb energy from gravitational waves (GWs) incident upon it, resulting in wave attenuation. This could in turn have implications for the interpretation of future data from early universe GWs. 
}

\section{Introduction} 

GWs emanating from the inflationary era of the cosmos en route to us
must have also traveled through plasma before the decoupling era. 
We attempt to show here that these GWs traveling through clouds of charged particles could suffer attenuation. The strain
satisfies 
 $$
 \Box^2h^{\mu\nu}  = - 2 \kappa T^{\mu\nu}   = 0; 
 \;\;\; h<\!\!\!<1  \eqno(1)
 $$
with  plane wave solutions in \emph{empty space} given by
 $$
 h^{\mu\nu} = A^{\mu\nu} exp(i \kappa_\rho x^\rho)\;\;\;
 \hbox{with}\;\;\;
 A^{\mu\nu}\kappa_\nu = 0
 $$
These waves are commonly thought of as having little or no interaction with matter, reaching us unattenuated, making GWs indispensable in detecting events during and after the inflationary era. \\

 A gauge in general relativity refers to a specific observer. In the TT gauge, the observer is exactly on a particle interacting with a GW, and for a wave
  propagating in the $\hat z$ direction, we have
 
 $$
 [h_{\mu\nu}^{TT}(t, z)] = \left(
 \begin{array}{cc}
 h_+ & h_\times \\
 h_\times & - h_+
 \end{array}
 \right)
 \cos\left[  \omega (  t -  \frac{z}{c} ) \right]
 $$
 In the TT gauge the coordinate positions of free particles remain unchanged since $  h_{00}, h_{0i} \rightarrow 0 $ due to the imposed gauge condition. In addition $\Gamma_{00}^i = 0$, which is only valid if $h^2$ terms are neglected. 
Despite this, in the proper detector frame, 
 particles will move with respect to one another.
 On Earth, the TT gauge is approximately the same as the 
 proper detector frame since 
 $\Gamma^i_{00} \approx 0  \; \; \hbox{and  }\;\; 
 h \sim 10^{-21}$ (expected)
 in our frame. 
 The motion of free particles in the proper detector frame can be described by a \emph{Newtonian type of force}
$$
F_i = \frac{1}{2} m\;   \ddot h_{ij}^{TT} \xi^j
$$

Therefore, we have the ability to describe the effects of GWs on particles in the proper
detector frame purely in a Newtonian way, without referring  to GR, remembering that
$h$ and $\xi$ are the values measured in the TT gauge \cite{maggiore08}!
However, this is possible if and only if Eq (1) is satisfied; that is, 
matter at the observing point does not add significantly to the 
curvature of spacetime. 
However, to use the geodesic equation to first order, we need to make sure that
the size of the detector is smaller than the wavelength of the GW.

\section{Interaction of GWs with Charged Particles}

For an astrophysical source, the GW strain at a distance $r$ is 
$
h \sim   
 2 G M v^2 /  r c^4  
$. 
To understand how such a wave would interact with a cloud of charged particles of appropriately small size, we need to look at the Einstein field equations \cite{herm}

$$
R_{\mu\nu}  - \frac{1}{2} \gmunu R = - \kappa  (  T_{\mu\nu} +  F_{\mu\nu} [EM]  ) \eqno(2)
$$
with
$$
g^{\mu\nu} \nabla_\mu \nabla_\mu  A_\alpha  = \mu_0 j_\alpha
\eqno(3)
$$
In addition, the following  Lorenz gauge condition should also be met. 
$$\;\;
\nabla_\alpha A^\alpha = 0
\;\;
\hbox{and} 
\;\;
\left[ A^\mu \right] = \left( \frac{\phi}{c} ,\;  \vec A \right)
$$

It is clear from Eq (1) that a TT gauge cannot be chosen inside 
a matter distribution as $T^{\mu\nu} \neq 0$. 
However, here  we assume that motionless (no currents are inside the cloud and $\vec B \neq 0$) charges do not 
contribute significantly to the curvature of spacetime. 
This can be assumed at least \emph{initially.}
Then (1) and (2) indicate that a  
\emph{TT frame, and thus a detector frame, 
 may be used even inside a charged cloud under the influence of a GW.} 
 Under these assumptions, Eq (3) gives the vector potential to be

$$
\Box^2 A_i = 0\;\;\; \hbox{(since there are no currents, at least initially)}
$$
and
$$
\Box^2 A_0 = c \mu_0 \rho +  h^{\mu\nu} \nabla_\mu\nabla_\nu A_0
\eqno(4)
$$

When no GWs are present solutions to the inhomogeneous 
wave equation (4), which satisfies 
the Lorenz gauge automatically, are well-known and are given by 

$$
A_0 = \frac{\mu_0}{4 \pi} \int d^3y\; 
 \frac{j_0( ct - |\vec x - \vec y|, \vec y    )}{|\vec x - \vec y |}
 =  \frac{1}{ 4 \pi \epsilon_0 c}  \int d^3y \frac{\rho }{|\vec x - \vec y |} \eqno(5) 
$$
where $\vec x$ is the field point while 
$\vec y$ is a point inside the charged distribution 
measured from the center of a spherical cloud. 
Integrals are evaluated at the retarded time $t_r = t - |\vec x - \vec y| / c$. 
Eq (5) gives  the zeroth order solution for $A_0$ in (4). A first order solution to (4) can be written as $A_0^{(1)} = \epsilon + A_0 $. 
Upon substitution into (4), this yields to first order (neglecting $\partial_\mu\partial_\nu \epsilon$)
$$
\Box^2\epsilon \approx  h^{\mu\nu} \partial_\mu\partial_\nu A_0
$$
Thus, the first order solution becomes 
$$
A_0^{(1)} =   
\frac{1}{ 4 \pi \epsilon_0 c}  \int d^3y \frac{\rho }{|\vec x - \vec y |}
 + \frac{1}{4 \pi} \int d^3 y \;
 \frac{ h^{\mu\nu} \partial_\mu\partial_\nu A_0}{ |\vec x - \vec y |  }\eqno(6) 
$$
While the first term of the foregoing 
gives the electrostatic potential due to charges,  
 the second term  emerging  from 
the derivatives of the charge density within the cloud  (see Eq (5)) results in a  time-varying vector potential giving 
rise to an electromagnetic (EM) field at the observation point $P$. 
Since this field is produced due to the GWs ($h^{\mu\nu}$) 
some amount of energy of the incident waves 
 is lost in transit via the cloud. 
\\

\noindent{\it GW of a Specific Simple Form}
\\

Consider the following specific GW form propagating 
along the $\hat r$ direction at a distance $r$ from 
a source, as an example.
$$
h^{\mu\nu} = \delta^\mu_\nu \; \frac{\lambda}{r} \cos \omega \left( t - \frac{r}{c} \right)
\;\;\; \hbox{with} \;\;\; h^{00} = h^{11} = 0 
$$
 Then, Eq (6) becomes 
 \begin{eqnarray*}
 A_0^{(1)} &=&   
  \frac{1}{ 4 \pi \epsilon_0 c}  \int d^3y \frac{\rho }{|\vec x - \vec y |} 
  +\frac{\lambda}{4 \pi r} \cos(\omega(t - r/c)) 
  \int d^3 y 
   \frac{  (\partial_z\partial_z A_0 - \partial_2\partial_2 A_0 )}{ |\vec x - \vec y |}\;\;\;\;\;\; (7)
 \end{eqnarray*}
 The EM fields may be easily calculated via 
 $$
 \vec E = - \nabla  c A^0   - \partial_t \vec A 
  \;\;\;  \hbox{and}\;\;\;
    B = \nabla \times \vec A
    \;\;\;
 \hbox{with}
 \;\;\;
 \partial_0 A^0 -  \nabla\cdot\vec A = 0
 $$

\section{Results and Discussion} 

We see from Eq (7) that in the asymptotic region far away from the 
charge distribution the Poynting flux is proportional to
$1/  |\vec x - \vec y |^2 $. 
Thus, the cloud of charges absorbs a \emph{finite amount}
of energy provided that those gradients of the charge density are nonzero. 
From Eq (7), we see that 
the amount of absorbed energy depends on the inhomogeneity 
of the charge cloud and how far it is from the source of GWs. 

For an EM wave, a charged particle oscillating in the transverse directions  stays at rest even after the wave passes showing that the particle does not absorb energy. 
This is not the case in the direction of propagation \cite{marsh11}. 
The displacement of the charged particle in the direction of 
propagation of the wave is proportional to
$1/2 (B_0^2 - A_0^2)$. 
Then it is clear that 
unless the incident radiation is circularly polarized, a charged 
particle will oscillate in the direction of propagation, absorbing energy. 

But for a GW, longitudinal proper oscillations in the direction 
of propagation of the wave will still not vanish even if the 
incoming GW is circularly polarized (  $h_+ = \pm h_\times$)
 \cite{marsh11}. 
Thus as time goes on, GWs sweeping through a charged cloud 
will modify the proper charge density and the gradients in 
Eq (7) will be established. 
The result is that the GW will be attenuated, losing energy absorbed by 
the charges in the cloud. 
If magnetic fields are present initially \cite{kleidis}, the attenuation will be even more pronounced as $A_i \neq 0$ in Eq (4). 

Our analysis shows that GWs incident upon an inhomogeneous 
cloud of charged particles will be attenuated even if 
the charges are not subjected to any magnetic field. 
Since waves change the proper charge density at least in the 
direction of the propagation of the wave, even a completely homogeneous 
cloud will eventually become inhomogeneous and the gradients 
of the charge density will be nonzero. 
When this occurs, we see from Eq (7) that an EM field will be 
produced and GW energy will be absorbed. It is clear that near a source, attenuation 
can be rather significant. 
Therefore, it is vital for us to consider these effects especially 
when trying to understand GWs from the inflationary era. 
This could have additional implications requiring a reevaluation of the 
magnitudes of $h$ from various sources expected here on Earth.

\end{document}